\documentclass[11pt,a4paper]{article}
\usepackage{graphicx}
\usepackage{amsmath}
\usepackage{cite}
\usepackage{amsfonts}
\usepackage{amssymb}
\usepackage[dvipdfm,hypertex,backref]{hyperref}
\newtheorem{theorem}{Theorem}

\newtheorem{lemma}{Lemma}

\newenvironment{proof}[1][Proof]{\textbf{#1.} }{\ \rule{0.5em}{0.5em}}
\begin{document}

\title{Synchronization of Networks with Prescribed Degree Distributions}
\author{Fatihcan M. Atay \thanks{Corresponding author, \texttt{atay@member.ams.org},
Tel: (+49) 341 9959-518, Fax: (+49) 341 9959-555}
\and T{\"{u}}rker B\i y\i ko\u{g}lu \thanks{\texttt{biyikogl@mis.mpg.de}}
\and J{\"{u}}rgen Jost \thanks{\texttt{jost@mis.mpg.de. }Also at Santa Fe
Institute, Santa Fe, NM 87501, USA; \texttt{jost@santafe.edu}} \medskip\\Max Planck Institute for Mathematics in the Sciences \\Inselstr. 22, D-04103 Leipzig, Germany}
\date{Preprint. Final version in \\
		 \emph{IEEE Transactions on Circuits and Systems I: \\
		   Fundamental Theory and Applications}}
\maketitle

\begin{abstract}%
We show that the degree distributions of graphs do not suffice to characterize
the synchronization of systems evolving on them. We prove that, for any given
degree sequence satisfying certain conditions, there exists a connected graph
having that degree sequence for which the first nontrivial eigenvalue of the
graph Laplacian is arbitrarily close to zero. Consequently, complex dynamical
systems defined on such graphs have poor synchronization properties. The
result holds under quite mild assumptions, and shows that there exists classes
of random, scale-free, regular, small-world, and other common network
architectures which impede synchronization. The proof is based on a
construction that also serves as an algorithm for building non-synchronizing
networks having a prescribed degree distribution.
\end{abstract}%

\bigskip


\section{Introduction}

Many network arhitectures encountered in nature or used in applications are
described by their degree distributions. In other words, if $P(d)$ denotes the
fraction of vertices having $d$ incident edges, then the shape of the function
$P(d)$ distinguishes certain network classes. For example, in the classical
random graphs studied by Erd\"{o}s and R\'{e}nyi \cite{Solomonoff51,Erdos59},
$P(d)$ has a binomial distribution, which converges to a Poisson distribution
for large network sizes. The degree distribution of regular networks, where
every vertex has the same degree $k$, are given by the delta function
$P(d)=\delta(d-k)$, whose small perturbations by random rewiring introduce
small-world effects \cite{Watts98,Newman-Watts99}. Many common networks, such
as the World-Wide Web \cite{Broder00}, the Internet \cite{Chen02}, and,
networks of protein interactions \cite{Jeong01} have been shown to have
approximate power-law degree distributions, and have been termed
\emph{scale-free} \cite{Barabasi99}. The power grid has exponentially
distributed vertex degrees \cite{Watts98}. Some social networks have
distributions similar to a power-law, possibly with some deviations or
truncations at the tails \cite{Grossman95}. A recent survey is given in
\cite{Newman03}. These examples demonstrate the recent widespread effort in
classifying common large networks according to their degree distributions.

In many applications, the vertices of networks have internal dynamics, and one
is interested in the time evolution of a dynamical system defined on the
network. A natural question is then what, if anything, the degree distribution
can say about the dynamics on the network. Asked differently, to what extent
does a classification according to degree distributions reflect itself in a
classification according to different qualitative dynamics? The question is
all the more significant since many large networks are constructed randomly.
Hence, in essence every realization of, say, a power-law distribution is a
different network. On the one hand, it is conceivable that the dynamics on
different realizations could be sufficiently different. On the other hand, it
is known that these realizations can have certain common characteristics; for
instance, small average distances and high local clustering found in
small-world networks \cite{Newman03}. Consequently, the relation between the
degree distribution and the dynamics is not trivial. The present work studies
this relation in the context of a specific but important dynamical behavior of
networked systems, namely synchronization.

It is well-known that the synchronization of diffusively-coupled systems on
networks is crucially affected by the network topology. In particular, the
so-called spectral gap, or the smallest nontrivial eigenvalue $\lambda_{1}$,
of a discrete Laplacian operator plays a decisive role for chaotic
synchronization: Larger values of $\lambda_{1}$ enable synchronization for a
wider range of parameter values, in both discrete and continuous-time systems
\cite{Jost02,Barahona02,Li-Chen03}, and also in the presence of transmission
delays \cite{Atay-PRL04}. Here we shall prove that, given any degree
distribution satisfying certain mild assumptions, a connected network having
that distribution can be constructed in such a way that $\lambda_{1}$ is
inversely proportional to the number of edges in the graph. Hence, there exist
large networks with these degree distributions which are arbitrarily poor
synchronizers. The proof is based on ideas from degree sequences of graph
theory, and in particular makes use of the relation between $\lambda_{1}$ and
the Cheeger constant of the graph. It applies to the class of distributions
that include the classical random (Poisson), exponential, and power-law
distributions, as well as nearest-neighbor-coupled networks and their
small-world variants, and many others. Thus, we establish that the degree
distribution of a network does not suffice to characterize the
synchronizability of complex dynamics on its vertices.

Our proof is constructive in nature. Hence, from another perspective, it can
be viewed as an algorithm for designing non-synchronizing networks having a
prescribed degree distribution. This may have implications for engineering
systems which should essentially operate in asynchrony, such as the Internet,
where synchronized client requests can cause congestion in the data traffic
through servers or routers, or neuronal networks where synchronization can be
sign of pathology. Furthermore, the non-synchronizing networks that we
construct can actually have quite small diameters or average distances. This
proves that the informal arguments which refer to efficient information
transmission via small distances as a mechanism for synchronization are
ill-founded. For a related discussion based on numerical studies of
small-world networks, see \cite{Barahona02,Nishikawa03}. Here we give a
rigorous proof for a much larger class of distributions.

\section{Degree sequences}

\label{sec:dist} In the following we consider finite graphs without loops or
multiple edges. As mentioned above, the degree distribution $P(d)$ gives the
fraction of vertices having $d$ incident edges. A related notion is that of a
\emph{degree sequence,} which is a list of nonnegative integers $\pi
=(d_{1},\ldots,d_{n})$ where $d_{i}$ is the degree of the $i$th vertex. We
also denote the largest and smallest degrees by $d_{\max}$ and $d_{\min}$,
respectively. For each graph such a list is well-defined, but not every list
of integers corresponds to a graph. A sequence $\pi=(d_{1},\ldots,d_{n})$ of
nonnegative integers is called \emph{graphic} if there exists a graph $G$ with
$n$ vertices for which $d_{1},\ldots,d_{n}$ are the degrees of its vertices.
$G$ is then referred to as a \emph{realization} of $\pi$. A characterization
of graphic degree sequences is given by the following.

\begin{lemma}
[\cite{Havel,Hakimi}]\label{Havel-Hakimi} For $n>1$, the nonnegative integer
sequence $\pi$ with $n$ elements is graphic if and only if $\pi^{\prime}$ is
graphic, where $\pi^{\prime}$ is the sequence with $n-1$ elements obtained
from $\pi$ by deleting its largest element $d_{\max}$ and subtracting 1 from
its $d_{\max}$ next largest elements. The only $1$-element graphic sequence is
$\pi_{1}=(0)$.
\end{lemma}

Often one is interested in connected graphs, and the following result gives
the correspondence between degree sequences and connected realizations (see
e.g. \cite{Melnikov}).

\begin{lemma}
\label{connected seq} A graphic sequence $\pi$ with $n$ elements has a
connected realization if and only if the smallest element of $\pi$ is positive
and the sum of the elements of $\pi$ is greater or equal than $2(n-1)$.
\end{lemma}

One of the simplest degree sequences belong to $k$-regular graphs, in which
each vertex has precisely $k$ neighbors. Here, the degree sequence is given by%
\begin{equation}
\pi=(k,\ldots,k). \label{reg}%
\end{equation}
There is a simple criterion for such constant sequences to be graphic (see
e.g. \cite{Melnikov}).

\begin{lemma}
\label{regular} A sequence $\pi=(k,\ldots,k)$ with $n$ elements where $k<n$ is
graphic if and only if $n$ or $k$ is even.
\end{lemma}

The construction of \emph{small-world} networks starts with a $k$-regular
graph with $k$ much smaller than the graph size $n$, e.g., a large circular
arrangement of vertices which are coupled to their near neighbors. Then a
small number of edges $c$ are added between randomly selected pairs of
vertices \cite{Newman-Watts99}. When $c\ll n$, the degree of a vertex
typically increases by at most one, which yields the degree sequence
\begin{equation}
\pi=(k+1,\dots,k+1,k,\dots,k) \label{sw1}%
\end{equation}
where $2c$ vertices have degree $k+1$. In a variant of the model
\cite{Watts98}, randomly selected edges are \emph{replaced} by others, so the
degree of a vertex may increase or decrease by one, and the degree sequence
becomes
\begin{equation}
\pi=(k+1,\dots,k+1,k,\dots,k,k-1,\dots,k-1) \label{sw2}%
\end{equation}
where the number of vertices having degree $k+1$ or $k-1$ are each equal to
$c$. More generally, there is also the possibility of having some vertex
degrees increase or decrease by more than one, in which case the sequences
(\ref{sw1}) and (\ref{sw2}) are modified accordingly, while the sum of the
degrees remain equal to $nk+2c$ and $nk$, respectively.

In addition to these well-known graph types, we shall also consider more
general sequences, which we define as follows.

\begin{description}
\item [ Definition ]A sequence $\pi$ with largest element $d_{\max}$ is called
a \emph{full sequence} if each integer $d$ satisfying $1\leq d\leq d_{\max}$
is an element of $\pi$, and the sum of the elements of $\pi$ is even.
\end{description}

We also give a criterion for full sequences to be graphic.

\begin{lemma}
\label{scale-free degreeseq} Let $\pi=(d_{\max},\ldots,d_{\max},d_{\max
}-1,\ldots,d_{\max}-1,\ldots,1,\ldots,1)$ be a full sequence with $n$ elements
and $d_{\max}\leq n/2$. Then $\pi$ is graphic.
\end{lemma}

\begin{proof}
We prove by induction on the number of elements of $\pi$. This is trivial for
the full sequence with two elements. Suppose that every full sequence with at
most $n\geq2$ elements and largest element not larger than $n/2$ is graphic.
Let $\pi$ be a full sequence with $n+1$ elements and largest element $d_{\max
}\leq(n+1)/2$. We look at the sequence $\pi^{\prime}$ that is defined in Lemma
\ref{Havel-Hakimi}. It is easy to see that $\pi^{\prime}$ is a full sequence.
Let $d_{\max}^{\prime}$ be the largest element of $\pi^{\prime}$. By the
definition of $\pi^{\prime}$, $d_{\max}^{\prime}\leq d_{\max}$. We claim that%
\begin{equation}
d_{\max}^{\prime}\leq n/2.\label{qq21}%
\end{equation}
For if $d_{\max}^{\prime}>n/2$, then%
\[
n/2<d_{\max}^{\prime}\leq d_{\max}\leq(n+1)/2.
\]
This implies
\begin{equation}
d_{\max}^{\prime}=d_{\max}=(n+1)/2\label{qq22}%
\end{equation}
so the number of $d_{\max}$'s in $\pi$ is at least $d_{\max}+2$. Since $\pi$
is a full sequence, the number of its elements is then%
\[
n+1\geq(d_{\max}+2)+(d_{\max}-1),
\]
implying $d_{\max}\leq n/2$, which contradicts (\ref{qq22}). Thus,
(\ref{qq21}) holds. Then $\pi^{\prime}$ is graphic by induction, and by Lemma
\ref{Havel-Hakimi} $\pi$ is graphic.
\end{proof}

The reason for introducing the concept of full sequences is that many common
graph types have full degree sequences. For instance, large random graphs of
Erd\H{o}s-R\'{e}nyi have their vertex degrees distributed according to the
Poisson distribution
\begin{equation}
P(d)\sim\frac{\mu^{d}e^{-\mu}}{d!},\quad d\geq1. \label{poisson}%
\end{equation}
Since such networks are randomly constructed, (\ref{poisson}) is understood to
hold in the limit as the network size increses while $\mu$ is kept constant.
In scale-free networks, the degree distribution follows a power law
\begin{equation}
P(d)\sim d^{-\beta} \label{power}%
\end{equation}
for some $\beta>1$. The exponential distribution obeys
\[
P(d)\sim e^{-\mu d}%
\]
with $\mu>0$. Of course, in any finite graph these distributions are truncated
from the right since the maximum degree is finite, and then one has $P(d)>0$
for $1\leq d\leq d_{\max}$. Furthermore, the sum of the vertex degrees is
always even since it is twice the number of edges in the graph. Therefore,
large finite graphs approximated by these distributions, or more generally by
any distribution satisfying $P(d)>0$ for $1\leq d\leq d_{\max}$, have
realizations with full degree sequences\footnote{In a particular realization
some degrees may actually be missing; we also address this possibility in
Section~\ref{sec:main}.}. In the next section, we prove that the regular and
small-world sequences (\ref{reg})-(\ref{sw2}), full degree sequences, as well
as their variants where some degrees may be missing, have a realization which
is a poor synchronizer. Furthermore, the synchronizability of this realization
worsens with increasing graph size.

\section{Constructing networks with small spectral gap}

\label{sec:main} The synchronization of dynamical systems is usually studied
in diffusively coupled equations, which in discrete-time may have the form
\begin{equation}
x_{i}(t+1)=f(x_{i}(t))+\kappa\left[  \frac{1}{d_{i}}\sum_{\substack{j\\j\sim
i}}f(x_{j}(t))-f(x_{i}(t))\right]  ,\;i=1,\dots,n.\label{coupled}%
\end{equation}
In (\ref{coupled}), $x_{i}$ denotes the state of the $i$th unit, which is
viewed as a vertex of a graph, and has $d_{i}$ neighbors to which it is
coupled. The notation $i\sim j$ denotes that units $i$ and $j$ are coupled,
which is represented in the underlying graph by an edge. Furthermore, $f$ is a
differentiable function and $\kappa\in\mathbf{R}$ quantifies the coupling
strength. A solution $(x_{1}(t),\dots,x_{n}(t))$ of (\ref{coupled}) is said to
synchronize if $\lim_{t\rightarrow\infty}\left|  x_{i}(t)-x_{j}(t)\right|  =0$
for all $i,j.$ The synchronization manifold $\mathcal{M}$ defined by the
conditions $x_{1}=\cdots=x_{n}$ is an invariant manifold for (\ref{coupled}),
and clearly the trajectories starting in $\mathcal{M}$ synchronize. In case
$\mathcal{M}$ is an attracting set, the system (\ref{coupled}) is said to
synchronize, and one distinguishes between different types of synchronization
depending on the nature of the attractor
\cite{Pikovsky-Grassberger91,Ashwin94,Hasler-Maistrenko97}. For example, when
$f$ is a chaotic map $\mathcal{M}$ may contain an attractor in the weak Milnor
sense \cite{Milnor85}, that is, for some set of initial conditions with
positive Lebesgue measure, and the corresponding behavior is then termed weak
synchronization. Systems similar to (\ref{coupled}) arise also in continuous
time in different applications. For a general introduction the reader is
referred to \cite{Pikovsky}.

The effect of the network topology on synchronization is determined by the
properties of the coupling operator. For (\ref{coupled}), the relevant
operator is the (normalized) Laplacian with matrix representation
\begin{equation}
L=D^{-1}A-I \label{Laplacian}%
\end{equation}
where $D=\mathrm{diag}\{d_{1},\dots,d_{n}\}$ is the diagonal matrix of vertex
degrees and $A=[a_{ij}]$ is the adjacency matrix of the graph, defined by
$a_{ij}=1$ if $i\sim j$ and zero otherwise. The eigenvalues of $L$ are real
and nonpositive, which we denote by $-\lambda_{i}$. Zero is always an
eigenvalue, and has multiplicity 1 for a connected graph. The smallest
nontrivial eigenvalue, denoted $\lambda_{1}$, is called the spectral gap of
the Laplacian, and is the important quantity for the synchronization of
coupled chaotic systems. Larger values of $\lambda_{1}$ enable chaotic
synchronization for a larger set of parameter values. This result holds in
both continuous and discrete time, for the Laplacian (\ref{Laplacian}) that we
consider here, as well as the combinatorial Laplacian $A-D$
\cite{Jost02,Barahona02,Li-Chen03}. In the following, we shall show how to
construct a graph having a prescribed degree sequence and an arbitrarily small
spectral gap, that is, a poor synchronizer.

Let $G=(E,V)$ be a connected graph, with edge set $E$ and vertex set $V$. We
denote the cardinality of a set $S$ by $|S|$. Thus $|E(G)|$ is the number of
edges of $G$. For a subset $S\subset V$, we define
\begin{equation}
h_{G}(S)=\frac{|E(S,V-S)|}{\min(\sum_{v\in S}d_{v},\sum_{u\in V-S}d_{u})},
\label{Cheeger}%
\end{equation}
where $|E(S,V-S)|$ denotes the number of edges with one endpoint in $S$ and
one endpoint in $V-S$. The \emph{Cheeger constant }$h_{G}$ is defined as
\cite[Section~2.2]{Chung97}
\[
h_{G}=\min_{S\subset V}h(S).
\]
This quantity provides an upper bound for the smallest nontrivial eigenvalue.

\begin{lemma}
\label{firsteig} $\lambda_{1}\leq2h_{G}$.
\end{lemma}

This result is basically Lemma~2.1 in \cite{Chung97}. There it is proved for
the smallest nontrivial eigenvalue of the matrix $\tilde{L}=[l_{ij}]$ defined
by%
\[
l_{ij}=\left\{
\begin{array}
[c]{cc}%
a_{ij}/\sqrt{d_{i}d_{j}} & \text{if }i\neq j\\
-1 & \text{if }i=j
\end{array}
\right.
\]
Since $\tilde{L}=D^{-1/2}AD^{-1/2}-I=D^{1/2}LD^{-1/2}$ is similar to the
matrix $L$ in (\ref{Laplacian}), its eigenvalues coincide with those of $L$.
Hence, the above lemma also holds for $L$.

Based on the estimate given in Lemma~\ref{firsteig}, we construct, from a
given degree distribution, a realization whose spectral gap $\lambda_{1}$ is
small. We first consider regular graphs.

\begin{theorem}
\label{thm:regular} Suppose $\pi=(k,\ldots,k)$ is graphic with $n$ elements,
with $2\leq k<n/2$. Then $\pi$ has a connected realization $G$ such that
\[
\lambda_{1}(G)\leq\frac{4}{|E(G)|-k}.
\]
\end{theorem}

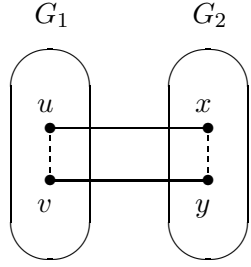
\begin{figure}[tb]
\begin{center}
\setlength{\unitlength}{0.7cm} \begin{picture}(8,5)
\put(2,2.5){\oval(1.5,4)}
\put(5,2.5){\oval(1.5,4)}
\put(2,2){\circle*{0.2}}
\put(2,3){\circle*{0.2}}
\put(5,2){\circle*{0.2}}
\put(5,3){\circle*{0.2}}
\put(2,2){\dashbox{0.1}(0,1)}
\put(5,2){\dashbox{0.1}(0,1)}
\put(2,2){\line(1,0){3}}
\put(2,3){\line(1,0){3}}
\put(1.7,5){$G_1$} \put(4.7,5){$G_2$}

\put(1.75,1.4){$v$} \put(1.75,3.3){$u$}

\put(4.75,1.4){$y$} \put(4.75,3.3){$x$}
\end{picture}
\end{center}
\caption{Partitioning the degree sequence $\pi$ in two parts $\pi_{1}$ and
$\pi_{2}$, where $G_{1}$ and $G_{2}$ are realizations of $\pi_{1}$ and
$\pi_{2}$, respectively. }%
\label{fig:1}%
\end{figure}

\begin{proof}
We split $\pi$ into two graphic sequences $\pi_{1}$ and $\pi_{2}$, which are
either equal (for $n$ equal 0 modulo 4 or $n$ equal 2 modulo 4 and $k$ is
even), or $\pi_{1}$ has one element less than $\pi_{2}$ (for $n$ equal 1 or 3
modulo 4), or $\pi_{1}$ has two elements less than $\pi_{2}$ (for $n$ equal 2
modulo 4 and $k$ is odd). (By splitting we mean that the union of the lists
$\pi_{1}$ and $\pi_{2}$ is equal to $\pi$.) By Lemmas \ref{connected seq} and
\ref{regular}, $\pi_{1}$ and $\pi_{2}$ have connected realizations $G_{1}$ and
$G_{2}$, respectively. Now we construct a connected realization $G$ of $\pi$
as follows (see Figure~\ref{fig:1}). Let $uv$ and $xy$ be edges of $G_{1}$ and
$G_{2}$ that are in a cycle. We delete $uv$ and $xy$, and add new edges $ux$
and $vy$. This new graph $G$ is connected, and it is a realization of $\pi$
since vertex degrees remain unchanged after this operation. Furthermore, if
$S$ denotes the smaller of the vertex sets $V(G_{1})$ and $V(G_{2})$, then
$(n-2)/2\leq|S|$ (see above), and we have
\begin{equation}
h_{G}(S)=\frac{2}{\sum_{v\in S}d_{v}}\leq\frac{2}{\frac{1}{2}nk-k}=\frac
{2}{|E(G)|-k}\label{qq31}%
\end{equation}
Thus by Lemma \ref{firsteig}, $\lambda_{1}(G)\leq4/(|E(G)|-k)$.
\end{proof}

Next we consider the small-world variants of regular graphs.

\begin{theorem}
\label{thm:sw} Let $\pi$ be a $k$-regular degree sequence with $n$ elements,
with $k<n/2$, and let $\pi^{\prime}$ be a small-world degree sequence obtained
from $\pi$ by adding or replacing $c$ edges. Then $\pi^{\prime}$ has a
connected realization $G$ such that
\[
\lambda_{1}(G)\leq\frac{2(c+2)}{|E(G)|-(k+c)}.
\]
\end{theorem}

\begin{proof}
First we split the $k$-regular degree sequence $\pi$ in two parts, as in the
previous proof, obtaining the situation shown in Figure~\ref{fig:1}, and
obtain the estimate (\ref{qq31}) for $h_{G}(S)$, where $S$ is the smaller of
the vertex sets $V(G_{1})$ and $V(G_{2})$. Now we add or replace $c$ edges in
$G$. The numerator in (\ref{qq31}) can then increase by at most $c,$ in case
the $c$ added adges are all between $G_{1}$ and $G_{2}$. Furthermore, the
denominator can decrease by at most $c$, in case all the removed edges, if
any, are all in $G_{1}$. Hence,
\[
h_{G}(S)\leq\frac{2+c}{|E(G)|-k-c}%
\]
and the result follows by Lemma~\ref{firsteig} as before.
\end{proof}

\begin{description}
\item [ Remark]By the same argument, a connected graph $G$ can be constructed
from a $k$-regular graph by \emph{removing} $c$ edges, whose spectral gap
satisfies%
\[
\lambda_{1}(G)\leq\frac{4}{|E(G)|-(k+c)}.
\]
\end{description}

The method of proof for the above results is actually applicable to a large
class of degree sequences, including the full sequences defined in
Section~\ref{sec:dist}.

\begin{theorem}
\label{thm:full} Let $\pi=(d_{\max},\ldots,d_{\max},d_{\max}-1,\ldots,d_{\max
}-1,\ldots,1,\ldots,1)$ be a full graphic sequence with $n$ elements,
$d_{\max}\leq n/4$, for which the sum of the elements is greater or equal than
$2n+d_{\max}$, and each $k$, $1\leq k\leq d_{\max}$, appears more than once.
Then $\pi$ has a connected realization $G$ such that
\[
\lambda_{1}(G)\leq\frac{4}{|E(G)|-d_{\max}/2}.
\]
\end{theorem}

\begin{proof}
For $k=1,\ldots,d_{\max},$ let $n_{k}$ be the number of times the element $k$
appears in the sequence $\pi$. We construct two sequences $\pi_{1}$ and
$\pi_{2}$ from $\pi$ as follows. The sequences $\pi_{1}$ and $\pi_{2}$ get
$\lfloor n_{k}/2\rfloor$ elements from each $k$ for $k=1,\ldots,d_{\max}$,
where the notation $\lfloor x\rfloor$ denotes the largest integer less than or
equal to $x$. Let $d_{1},\ldots,d_{s}$ be the remaining elements of $\pi$,
where the $d_{i}$ are distinct. For the set $N=\{1,2,\ldots,s\}$, we define
\[
r_{\pi}=\min_{J\subseteq N}\left|  \sum_{j\in J}d_{j}-\sum_{j\notin J}%
d_{j}\right|  .
\]
By induction on the number of elements, it is easy to see that $r_{\pi}\leq
d_{\max}.$ Now we spread the remaining elements $d_{1},\ldots,d_{s}$ to the
sequences $\pi_{1}$ and $\pi_{2}$ such that $r_{\pi}$ is minimum. By Lemmas
\ref{scale-free degreeseq} and \ref{connected seq}, $\pi_{1}$ and $\pi_{2}$
have connected realizations $G_{1}$ and $G_{2}$, respectively. Let $uv$ and
$xy$ be edges of $G_{1}$ and $G_{2}$ that are in a cycle or a vertex of the
edges has degree one. (Such edges exist since $G_{1}$ and $G_{2}$ are
connected.) We rewire these as before (Fig.~\ref{fig:1}), and the rest of the
proof follows analogously to the proofs of Theorems \ref{thm:regular} and
\ref{thm:sw}.
\end{proof}

We finally generalize to include degree sequences where some degrees between
$1$ and $d_{\max}$ may be missing.

\begin{theorem}
\label{thm:ext} Let $\pi=(d_{\max},\ldots,d_{\min})$ be a graphic sequence
with $n$ elements, where $d_{\max}\leq n/6$, $d_{\min}\geq2,$ and each degree
appears more than once. Let $c$ be the sum of distinct integers between 1 and
$d_{\max}$ which do not appear in $\pi$. Then $\pi$ has a connected
realization $G$ such that
\begin{equation}
\lambda_{1}(G)\leq\frac{2c}{|E(G)|-d_{\max}/2}.\label{mineig-ext}%
\end{equation}
\end{theorem}

\begin{proof}
First, we follow the proof of Theorem~\ref{thm:full} and construct two
sequences $\pi_{1}$ and $\pi_{2}$ from $\pi$ as before. The sequences $\pi
_{1}$ and $\pi_{2}$ get $\lfloor n_{k}/2\rfloor$ elements from each present
degree. We also handle the remaining elements of $\pi$ as before. So, the sum
of the elements of the sequences differ by at most $d_{\max}$. Notice that by
construction, the sequences $\pi_{1}$ and $\pi_{2}$ have at least $n/3$
elements. It is possible that $\pi_{1}$ and $\pi_{2}$ are not graphic. By the
following process, we make these sequences graphic by adding new vertices and
edges. For a nonnegative integer sequence $s=(d_{1},\ldots,d_{n})$ with at
least one positive element, let $d$ be the largest element of $s$ and $p$ be
the number of positive elements of $s$. If $d>p-1$, then we add $d-p+1$ new
elements to $s$ that are equal to one, otherwise the sequence $s$ does not
change. Now we obtain the sequence $s^{\prime}$ from $s$ by deleting its
largest element $d$ and subtracting 1 from its $d$ next largest elements. We
repeat this for $s^{\prime}$ and so forth until the considered sequence has no
positive elements. Let $z$ be the total number of new elements that are added
in each step. By Lemma~\ref{Havel-Hakimi}, the sequence $s$ is graphic by
adding at most $z$ edges. We call this process a \emph{lay-on process} and say
that the sequence $s$ \emph{is graphic with at most $z$ new edges}. In other
words, if we add to the sequence $s=(d_{1},\ldots,d_{n})$ $z$ new elements
that are equal one, then the new sequence $s^{\ast}=(d_{1},\ldots
,d_{n},1\ldots,1)$ is a degree sequence. We call $s^{\ast}$ as
\emph{extension} of $s$. Since we can add all the missing elements of
$d_{\max},\ldots,1$ to $\pi_{1}$ and $\pi_{2}$ and by Lemma~\ref{scale-free
degreeseq} these new sequences are graphic, it follows that the sequences
$\pi_{1}$ and $\pi_{2}$ are graphic with at most $c$ new edges. \newline
\textbf{Claim 1:} If the sum of the elements of $s$ is even (odd), then $z$ is
even (odd). \newline The proof of the claim follows by induction on the number
of elements of $s$ and the lay-on process. \newline The sums of the elements
of $\pi_{1}$ and $\pi_{2}$ are either both even or both odd, because the sum
of the elements of $\pi$ is even. Then the total number of new elements of
$\pi_{1}$ and $\pi_{2}$ have the same parity, by the above Claim. We say a
sequence $s=(d_{1},\ldots,d_{n})$ has a connected \emph{extension}, if its
extension $s^{\ast}$ has a connected realization. \newline \textbf{Claim 2:}
The sequence $s=(d_{1},\ldots,d_{n}),$ $d_{1}\geq\cdots\geq d_{n}\geq2$, has a
connected extension. \newline We prove the claim by induction on the number of
elements of $s$. If $s$ has one element, it is trivial. We assume that the
claim is true, if $s$ has less than $n$ elements. Let $s$ have $n$ elements.
If $d_{n}\geq3$, then $s$ has a connected extension, by induction and the
lay-on process. Suppose now that $d_{n}=2$. If $d_{1}\geq d_{2}\geq3$, then we
delete $d_{n}$ and subtract one from $d_{1}$ and $d_{2}$, obtaining a sequence
with $n-1$ elements that are greater than one. By induction this sequence has
a connected extension. It follows that $s$ has a connected extension. It
remains to consider the case when $d_{1}>2$ and all other elements are equal
to two. For $d_{1}\leq n-4$, there is a vertex $v_{1}$ with $d_{1}$ neighbors
and the rest of the vertices can build up a cycle. We switch an edge of the
cycle and an edge of $v_{1}$, obtaining a connected extension of $s$. If
$d_{1}\geq n-1$, then by construction $s$ has connected extension. If
$d_{1}=n-3$, then there are two remaining vertices with degree two. By
construction, there is an edge between them and both of them have an edge with
neighbors of $v_{1}$. It follows that $s$ has a connected extension. The case
$d_{1}=n-2$ is similar. \newline The sequences $\pi_{1}$ and $\pi_{2}$ are
graphic with at most $c$ new edges, they have the same parity, and, by Claim
2, $\pi_{1}$ and $\pi_{2}$ have connected extension. It remains to construct a
connected realization of the degree sequence $\pi$. Let $G_{1}$ and $G_{2}$ be
the connected realizations of the extensions $\pi_{1}^{\ast}$ and $\pi
_{2}^{\ast}$, respectively. For the vertices $v_{1}$ and $v_{2}$ of $G_{1}$
and $G_{2}$ that are adjacent to the vertices with degree one, respectively.
We delete these edges and add an edge between $v_{1}$ and $v_{2}$, therefore
we get a connected realization $G$ of $\pi$. Assume that there remain in
$G_{1}$ two edges $ux_{1}$ and $vx_{2}$, where $x_{i}$ have degree one
(because of the same parity this case is the only possibility). Then we delete
these edges, as well as an edge $ab$ of $G_{2}$, and add the edges $au$ and
$bv$. Notice that after deleting and adding edges, the new graph is also
connected. $G$ can be split in two parts with at most $c$ edges between them,
and the sum of the degrees of these two parts differ by at most $d_{\max}$.
The estimate (\ref{mineig-ext}) follows as before from (\ref{Cheeger}) and
Lemma~\ref{firsteig}.
\end{proof}

\begin{description}
\item [ Remark]The upper bound for $\lambda_{1}$ given in
Theorem~\ref{thm:ext} is proportional to the sum of the missing degrees $c$;
hence, it is smaller if the degree sequence is closer to being a full sequence.
\end{description}

We note that many of the assumptions in Theorems \ref{thm:regular}%
-\ref{thm:ext} can be relaxed. Furthermore, the estimates obtained for
$\lambda_{1}$ are certainly not tight. Our aim here is not to obtain estimates
in full generality, but rather show that the spectral gap cannot be determined
from the degree distribution. Hence, we content ourselves to constructing
connected graphs for which $\lambda_{1}$ is inversely proportional to the
number of edges, i.e. is arbitrarily small for large graph sizes. Thus, based
on Theorems \ref{thm:regular}-\ref{thm:ext}, we conclude that the degree
distribution of a network in general is not sufficient to determine its synchronizability.

The construction used in the above proofs has a quite general nature. In
essence, the idea is to split a given sequence $\pi$ into two sequences
$\pi_{1},\pi_{2}$ more or less equal in size, which have the connected
realizations $G_{1},G_{2}$, respectively. Then, as schematically shown in
Figure~\ref{fig:1}, switching the edges between two pairs of vertices yields a
connected graph $G$ as a realization of $\pi$. The spectral gap $\lambda_{1}$
is estimated from the Cheeger constant of the sets $G_{1}$ or $G_{2}$, and is
small since the numerator in (\ref{Cheeger}) is 2 and the denominator can be
rather large. This gives an algorithm for constructing networks having a
prescribed degree distribution and a small spectral gap. In certain cases, it
may require some effort to obtain connected realizations for the partitions
$\pi_{1}$,$\pi_{2}$. Nevertheless, by arguments similar to the above, it is
not hard to show that if $\pi_{1}$ has a realization with $m$ components, then
it is possible to construct a realization $G$ of $\pi$ such that $\lambda_{1}$
is essentially given by $\lambda_{1}\sim2m/|E(G)|$. So, the basic construction
is indeed applicable to even more general cases and distributions than we have
considered here. On the other hand, it should be kept in mind that certain
degree sequences are not amenable to such manipulation, in particular those
that have a unique representation. For instance, the $n$-element sequences
$(n-1,\dots,n-1)$ and $(1,1,\dots,1,n-1)$ correspond uniquely to the complete
graph $K_{n}$ and the star $S_{n}$, respectively; hence the value of
$\lambda_{1}$ is uniquely determined for them.

We conclude this section with an example illustrating the main ideas.

\begin{figure}[tbh]
\begin{center}
\includegraphics[]{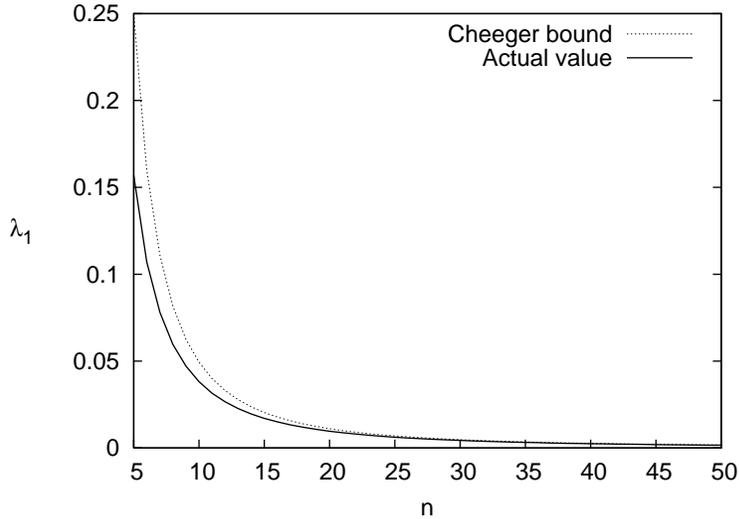}
\end{center}
\caption{The estimated and actual values of $\lambda_{1}$ for the graph
constructed in Example.}%
\label{fig:2}%
\end{figure}

\begin{description}
\item [ Example. ]Consider a constant degree sequence $\pi=(n-1,\dots,n-1)$ of
$2n$ elements, where $n\geq3$. We split it into two constant sequences
$\pi_{1}=\pi_{2}=(n-1,\dots,n-1)$ each with $n$ elements. Since either $n$ or
$n-1$ is even, both subsequences are graphic by Lemma~\ref{regular} and have
connected realizations $G_{1},G_{2}$ by Lemma~\ref{connected seq}. In fact, in
this special case the realization is unique: $G_{1}=G_{2}=K_{n}$, the complete
graph on $n$ vertices where each vertex is connected to the remaining $n-1$.
We now remove one edge each from $G_{1}$ and $G_{2}$, and rewire to construct
the structure shown in Figure~\ref{fig:1}, i.e., a connected graph $G$ having
the original degree sequence $\pi$. The subgraphs $G_{1},G_{2}$ are complete
graphs each missing an edge. To show that $G$ has a small spectral gap,
Theorem~\ref{thm:regular} can be invoked to obtain the bound%
\begin{equation}
\lambda_{1}\leq\frac{4}{(n-1)^{2}}\label{est}%
\end{equation}
which is plotted in Figure~\ref{fig:2} against $n$, together with the true
values of $\lambda_{1}$. The figure shows that $\lambda_{1}$ decreases rapidly
with increasing graph size, and is in fact estimated very well by the upper
bound (\ref{est}) for large $n.$ In this special case the estimate (\ref{est})
can actually be somewhat improved, but the point is that the value of
$\lambda_{1}$ is on the order of $n^{-2}$ for large $n$. It is worth noting
that the average distance in the constructed graph $G$ is small and approaches
2 as $n\rightarrow\infty$, and the maximum distance between any two vertices
is equal to 3. Despite these small distances, $G$ is a poor synchronizer even
at moderate sizes $n$.
\end{description}

\section{Discussion and conclusion}

We have shown that the degree sequence is generally not sufficient to
characterize the synchronizability of a network. The method of proof is based
on the construction of a graph which has a specified degree distribution and a
small spectral gap for the graph Laplacian. The construction works for a wide
class of degree distributions, including Poisson, exponential, and power-law
distributions, regular and small-world networks, and many others. Thus,
sychronizability is not an intrinsic property of a degree distribution.
Furthermore, small diameters or average distances in a graph do not
necessarily imply synchronization, since the poorly-synchronizing graphs we
construct can have small diameters, as the above Example shows.

The method of proof presented here is essentially an algorithm to construct a
graph with a given degree distribution and a small spectral gap. Such a
poor-synchronizing realization of a degree distribution is certainly not
unique. An interesting query is then finding how many poor-synchronizers there
are, as compared to all realizations of a given distribution? This turns out
to be a difficult question. In general it is not easy to estimate of the
number of realizations of a given degree sequence. McKay and Wormald
\cite{McKay91} give an estimate that depends on the maximum degree. To state
their result, let $\pi=(d_{1},\ldots,d_{n})$ be a degree sequence without zero
elements, $n_{1}$ entries of value $1$, and $n_{2}$ entries of value $2$. If
$n_{1}=O(n^{1/3}),$ $n_{2}=O(n^{2/3}),$ and $d_{\max}\leq\frac{1}{3}\log
n/\log\log n$, then the number of the realizations of $\pi$ is asymptotically%
\[
\frac{M!}{(M/2)!2^{M/2}d_{1}!\cdots d_{n}!n!}\exp\left(  -O(d_{\max}%
^{6})-O\left(  \frac{d_{\max}^{10}}{12n}\right)  +O\left(  \frac{d_{\max}^{3}%
}{M}\right)  \right)  ,
\]
where $M=\sum d_{i}=2|E(G)|$ \cite{McKay91}.

As an attempt to apply this result to the case considered here, let
$\pi=(d_{1},\ldots,d_{2n})$ be a full degree sequence with $2n$ elements and
$\pi_{1}=(d_{1},\ldots,d_{n})$ be one of the partitions with $n$ elements such
that $\sum_{i=1}^{2n}d_{i}=2\sum_{i=1}^{n}d_{i}$. Then the ratio of the number
of realizations $\pi_{1}$ to $\pi$ is approximately (by using $n!\approx
\sqrt{2\pi n}(\frac{n}{e})^{n})$
\[
d_{n+1}!\cdots d_{2n}!n^{n}2^{-M+2n}M^{-M/2}\exp\left(  M/2-n-O\left(
\frac{d_{\max}^{10}}{12n}\right)  \right)  ,
\]
where $M=\sum_{i=1}^{n}d_{i}=O(n\log n/\log\log n)$. For a certain partition
$\pi_{1}$ the ratio is very small. On the other hand it is hard to give an
estimate on the number of suitable partitions like $\pi_{1}$ that make
$\lambda_{1}$ very small. Hence, at this point there seems to be no general
estimates for the relative size of poor-synchronizers belonging to the same
degree sequence, and the question remains open for future research.

At any rate, it is clear that care is needed when making general statements
about the synchronizability of, say, scale-free networks, even in a
statistical or asymptotical sense. In this context it is also worth noting
that most results about the scale-free architecture are based on the algorithm
of Barab\`{a}si and Albert \cite{Barabasi-Albert99}, which, by way of
construction, possibly introduces more structure to affect the dynamics than
is reflected in its power-law degree distribution. In fact, another
preferential attachment type algorithm introduced in \cite{Jost-Joy02b} yields
also graphs with power-law degree distributions, but significantly smaller
first eigenvalues than those constructed by the algorithm of Barab\`{a}si and
Albert or random graphs. While the algorithm of \cite{Barabasi-Albert99} lets
nodes receive new connections with a probability proportional to the number of
connections they already possess, the one of \cite{Jost-Joy02b} introduces new
nodes by first connecting to an arbitrary node and then adding further
connections that complete triangles, that is, to nearest neighbors of nodes
already connected to. Since the probability to be a nearest node of another
node is also proportional to the number of links a nodes possesses, that
algorithm also implements a preferential attachment scheme and therefore
produces a power-law graph. The small first eigenvalues reported in
\cite{Jost-Joy02b} agree with our more general results in showing that it is
possible to construct power-law graphs with arbitrarily small spectral gap.
Furthermore, this observation holds regardless of the exponent in the power law.

\newpage

\end{document}